\documentstyle[epsf, rotate]{elsart}

\begin{document}
%
%
%
%
 
\begin{frontmatter}
  \title{Tsallis statistics with normalized q-expectation values is
    thermodynamically stable: illustrations} \author{A.  R.
    Lima\thanksref{email}} \thanks[email]{e-mail: arlima@if.uff.br}
  \address{Instituto de F\'{\i}sica, Universidade Federal Fluminense Av.
    Litor\^anea, s/n$^o$ - 24210-340 Niter\'oi, RJ, Brazil} \author{T.  J.  P.
    Penna\thanksref{email2}} \thanks[email2]{e-mail: tjpp@if.uff.br}
  \address{Instituto de F\'{\i}sica, Universidade Federal Fluminense Av.
    Litor\^anea, s/n$^o$ - 24210-340 Niter\'oi, RJ, Brazil} \address{Center
    for Polymer Studies, Boston University, 02215 Boston, MA, USA}

\begin{abstract}
  We present a study of both the ``Iterative Procedure'' and the ``$\beta
  \rightarrow \beta'$ transformation'', proposed by Tsallis et al ({\it
    Physica {\bf A261}, 534}) to find the probabilities $p_i$ of a system to
  be in a state with energy $\epsilon_i$, within the framework of a
  generalized statistical mechanics.  Using stability and convexity arguments,
  we argue that the iterative procedure does not always provide the right
  temperature dependence of thermodynamic observables.  In addition, we show
  how to get the correct answers from the ``$\beta \rightarrow \beta'$
  transformation''.  Our results provide an evidence that the Tsallis
  statistics with normalized q-expectation values is stable for all ranges of
  temperatures.  We also show that the cut-off in the computation of
  probabilities is required to achieve the stable solutions.
\end{abstract}


\end{frontmatter}

The study of alternatives to the Boltzmann-Gibbs statistics has arisen
increasing interest on the last years. Among several proposals, the most
widely used and studied has been the so-called ``Tsallis
statistics''\cite{tsallis88}.  This new formulation seems to be appropriate
for the study of non-extensive systems \cite{tsallispage}.  The list of
non-extensive systems, and that are reported as candidates to obey the Tsallis
statistics, is growing fast.  Therefore, the understanding of the basic
principles and properties of this non-extensive statistics has gain
fundamental importance. In this direction a recent work by Tsallis, Mendes and
Plastino \cite{tsallis98} covers the role of constraints within the Tsallis
statistics.  In that work, they study three different alternatives for the
internal energy constraint.  The first two choices correspond to the ones
which have been applied to many different systems in the last years
\cite{tsallispage}.  They are: (ia) $\sum_i{p_i \epsilon_i}=U$ and (ib)
$\sum_i{p_i^q \epsilon_i}=U_q$.  However, both constraints present undesirable
difficulties.  A third choice for the internal energy constraint is considered
by them \cite{tsallis98},

\begin{equation}
\label{newcons}
U_q=\frac{\sum_{i=1}^{\Omega}{p_{i}^{q}\epsilon_i}}
{\sum_{i=1}^{\Omega}{p_{i}^{q}}}, \label{constrain}
\end{equation}

where $q$ comes from the entropy definition \cite{tsallis88}

\begin{equation}
S_q=k\frac{1-\sum_{i=1}^{\Omega}{p_i^q}}{q-1},\label{sq}
\end{equation}

with $\sum_i{p_i}   =1 $,  where  $i$ is  a  given  state  with energy
$\epsilon_i$ from $\Omega$  possible states. Each constraint (ia),(ib)
and eq.(\ref{constrain}) determines a different set of  probabilities $p_i$ for
each   state with energy   $\epsilon_i$.    The extremization of   the
generalized entropy (\ref{sq}),  under constraint  \ref{constrain}
gives us an implicit equation for the probabilities $p_i$:

\begin{equation}
p_i=\left[1-\frac{(1-q)\beta(\epsilon_i-U_q)}
{\sum_{j=1}^{\Omega}{p_j^{q}}}\right]^{\frac{1}{1-q}}/Z_q\label{piq} 
\end{equation}

with 

\begin{equation}
Z_q(\beta)               \equiv                    \sum_{i=1}^{\Omega}
\left[1-\frac{(1-q)\beta(\epsilon_i-U_q)}
{\sum_{j=1}^{\Omega}{(p_j)^{q}}}\right]^{\frac{1}{1-q}} 
\end{equation}

The    normalized q-expectation value   of  an observable is therefore
defined as 

\begin{equation}
O_q             \equiv       \langle      O_i    \rangle_q      \equiv
\frac{\sum_{i=1}^{\Omega}{p_i^q O_i}}{\sum_{i=1}^{\Omega}{p_i^q}} 
\end{equation}

where  $O$  is any  observable   which commutes with  the Hamiltonian - 
otherwise we should  use make use of  the density operator $\rho$.  We
will refer to  this reformulation of  the Tsallis statistics as ``with
normalized q-expectation values''. 

In order to solve equation \ref{piq} Tsallis {\it et al.}  suggest two
different  approaches, namely the {\em   Iterative Procedure} and the {\em
$\beta  \rightarrow  \beta'$    transformation}.  In   the   iterative
procedure,  we start with an  initial set of  probabilities and iterate
them  self-consistently until the desirable  precision is reached.  In
the $\beta  \rightarrow  \beta'$ transformation the set   of equations
above is transformed to: 

\begin{equation}
p_{i} =  \left[1-(1-q)\beta'\epsilon_{i}\right]^{\frac{1}{1-q}}/Z'_{q}
\\ 
\end{equation}

\begin{equation}
Z'_{q}\equiv
\sum_{j=1}^{\Omega}{\left[1-(1-q)\beta'\epsilon_j\right]^
{\frac{1}{1-q}}}\\
\end{equation}

with

\begin{equation}
\label{betabetaast}
\beta'(\beta)\equiv             \frac{\beta}{              {(1-q)\beta
U_{q}+\sum_{j=1}^{\Omega} p_j^q}}.
\end{equation}

In order to obtain ${p_i}$, we go through the following steps:
\begin{enumerate}
\item Compute  the quantities   $y_i=(1-(1-q)\beta'\epsilon_{i}), \;\;
\forall i\in\Omega $.
\item If $y_i<0$ them $y_i=0$;
\item Compute $Z_{q'}=\sum_{i=1}^{\Omega}{y_i^{1/(1-q)}}$;
\item Compute $p_i(\beta')=y_i^{1/(1-q)}/Z_{q'}$;
\item Obtain  $U_q(\beta')$ and any  other thermodynamical  quantities
using equation 5;
\item Obtain $\beta(\beta')$ from equation (\ref{betabetaast}).
\end{enumerate}

This  recipe   allows  the determination    of   $p_i(\beta)$ for  all
$\beta(\beta')$  and consequently    $U_q(\beta)$   (and   any   other
observable). The second step in the above  procedure is the well known
(and useful)  cut-off \cite{tsallis98} associated  to {\em ``vanishing
probabilities''}.  This cut-off is  needed only for $q<1$. Because the
cut-off is  applied before the  computation of the  probabilities, the
norm constraint is still respected.

In ref.\cite{tsallis98},  the   authors  illustrate  their ideas    by
applying them to   a  system  with  discrete  spectrum   $\epsilon_n=n
\epsilon$ with $\epsilon > 0$ and $n=0,1,2,..., N$.  For $N=1$ we have
the  case  of two non-degenerate    levels.  The limit  $N \rightarrow
\infty$ corresponds to the quantum harmonic oscillator. Fig. 3 in that
paper presents the results for the internal energy $u_q=U_q/\epsilon$,
as  a function of the  temperature,  for both procedures.  The  $\beta
\rightarrow \beta'$ transformation  generates a non-physical reentrant
behavior whereas  the iterative  procedure produces a discontinuity
in  the free energy.  These  results suggest that  the system might be
unstable.  Therefore, we need to  figure out the relation between both
approaches.  We  show, through a simple example,  how to get rid of the
reentrant  region and consequently  that  the Tsallis statistics  with
q-normalized expectation values is  stable for all $T$.  This evidence
is  specially important  because theoretical arguments  for stability
similar  to the  ones  previously used  \cite{ramshaw95,tsallis95} are
very difficult (if  at all possible)  to be obtained from the implicit
equation (\ref{piq}).

We consider the same system as Tsallis {\it et al.}: a system with discrete
spectrum $\epsilon_n=n \epsilon$ with $\epsilon > 0$ and $n=0,1,2,..., N$.  We
focus on $q<1$ since for $q \geq 1$ no stability problems occur.  We start by
computing the generalized free energy, $F_q=U_q-\frac{1}{\beta}\ln_q Z_q$,
where $\ln_q(x)= (x^{1-q}-1)/(1-q)$, for a two level system.  In figure (1) we
present the free energy $f \equiv F_q/\epsilon$, as obtained from both
procedures, as a function of the normalized temperature $t=T/\epsilon$.

\begin{figure}[!ht]
\epsfxsize=8cm \centerline{ \rotate[r]{\epsfbox{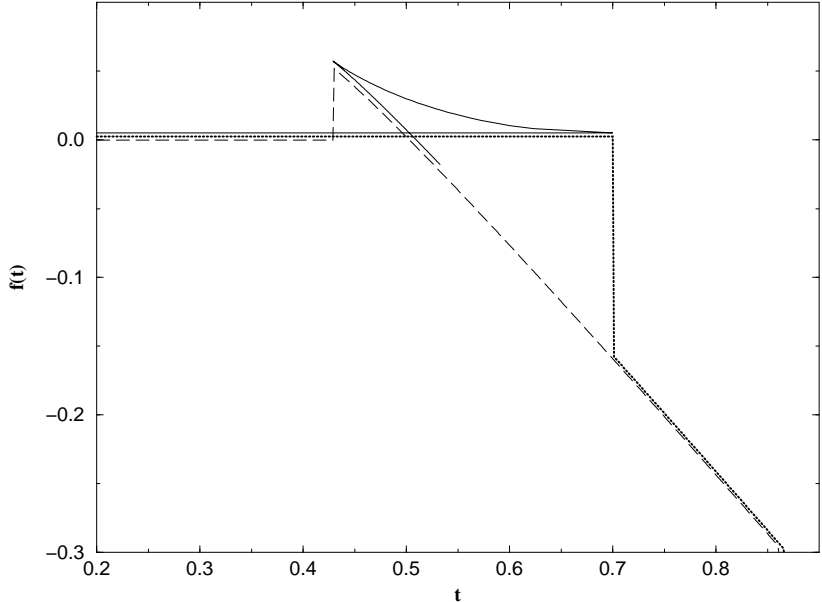}}}
\caption{
  Free energy calculated for the two-level system with $q=0.3$.  The curves
  are slightly displaced up only for the purpose of better visualization.  The
  results from the $\beta\rightarrow\beta'$ transformation (solid line) show
  us an unphysical ``loop''.  There are two possible paths for the iterative
  procedure: the dashed line is obtained by starting from a $t=0$
  configuration.  If the initial configuration is the one at $t=\infty$, the
  system follows the doted line.  Outside the region shown in this figure,
  both results give the same results.}
\end{figure}

Figure (1) shows us  that the iterative procedure  gives a free energy
curve with a jump. The jumps  will appear at different temperatures if
we take different initial conditions due to a feature of the iterative
method of solving non-linear equations: the system  tends to go to the
nearest solution.  The states that are reached for only one of the two
paths shown in figure (1) should correspond to metastable states.  We
should consider only  the one  with  the lowest free energy.   In this
way, we get rid of the uncomfortable  jumps: the free energy cannot be
discontinuous since  its convexity   must be  guaranteed for  all  $T$
\cite{stanley71}.  Because  of  the simplicity  of the  system treated
here (an energy spectrum  with only two levels),  it  is easy  to find
both paths.  However, as we are going to show in this work, it is even
easier  to get the  correct behavior from the $\beta\rightarrow\beta'$
transformation.

The  curve  for the  free    energy from the  $\beta\rightarrow\beta'$
transformation   displays a closed  loop.  This   loop deserves also a
careful analysis.   The  curved line must  be discarded  by  convexity
arguments (the specific heat would  be negative \cite{stanley71}). The
remaining states are the metastable states  (the same states that were
obtained from the iterative procedure).  Nevertheless, we can also get
rid of the loop by choosing always the states  with the lowest energy.
Similar procedures and  discussions can be found  in \cite{figueira94}
and \cite{callen}.  In  figure 2, we show a  closer look on the loops,
for  some values of   $q$.   For this system,    we find no  reentrant
behavior for $q\ge 0.56$.

\begin{figure}[!ht] 
\epsfxsize=8cm 
\centerline{\rotate[r]{\epsfbox{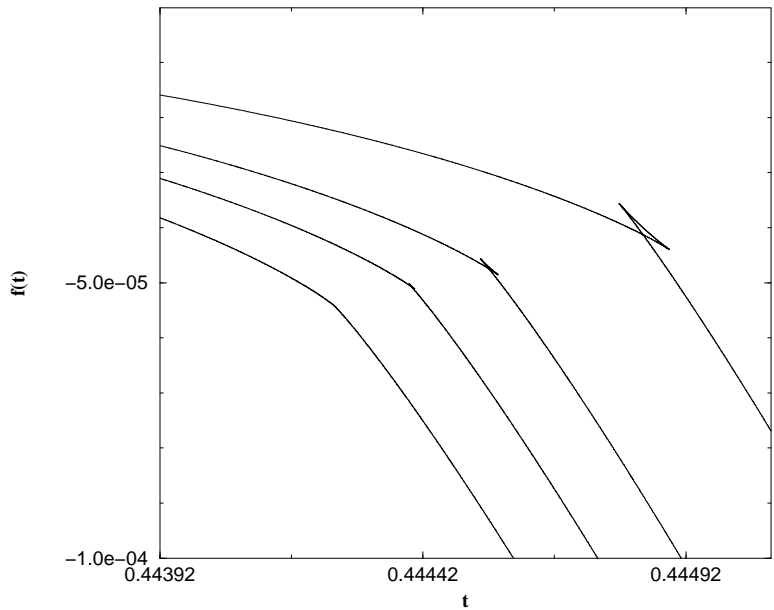}}}
\caption{
Closer  look on the  loops  on the free  energy  vs. temperature.  The
curves  from     the    right  to      the  left     corresponds    to
$q=0.558,0.5595,0.55975,0.56$.  The  loop disappears as $q$ approaches
$0.56$.}
\end{figure}

The  procedure  we have proposed  (to  look first at  the  free energy
curve)  gives  the correct behavior    for the   internal energy   and
consequently  for   the   computation   of  the other  thermodynamical
quantities  besides explaining  why the  iterative  procedure and  the
$\beta\rightarrow\beta'$ predict different  behaviors for the internal
energy.   In fig. 3 we show  the corrected curves for the internal
energy as function  of the temperature.   For the sake of  clarity, we
compare   this new     recipe   with  the     $\beta\rightarrow\beta'$
transformation.   For $q>0.56$ we  do  not need to  apply  any further
correction because there are no loops on  the generalized free energy.
As shown by Tsallis {\it et al.} \cite{tsallis98}, for those values of
$q$ both procedures give the same results.  The correct expression of
the temperature transformation for a  $N=8$-level system is shown  in
fig. 4.  We decided to show this result for  a richer system then for
the simple  two-level system,  in order to  emphasize the  role of the
spurious     states   introduced  by  the     $\beta\rightarrow\beta'$
transformation.

\begin{figure}[!ht]

\epsfxsize=8cm \centerline{ \rotate[r]{\epsfbox{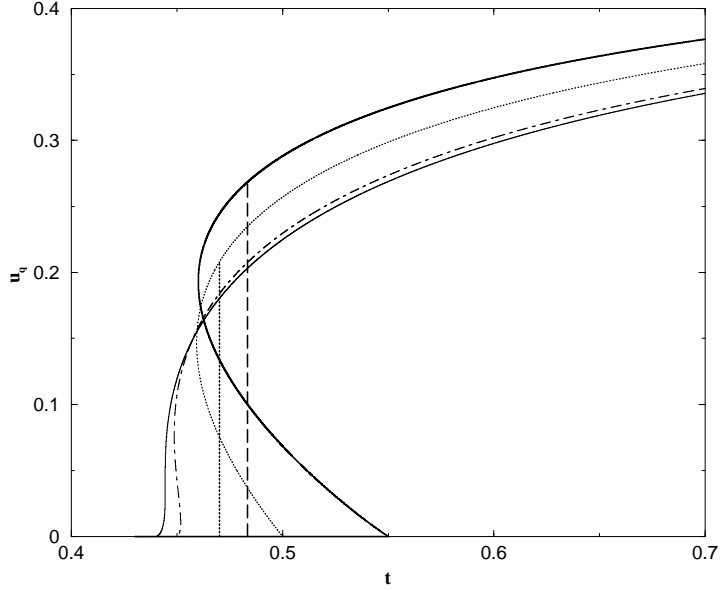}}}
\caption{ 
  Internal energy as function of the normalized temperature.  The thick solid
  line correspond to $q=0.45$ internal energy obtained from a
  $\beta\rightarrow\beta'$ transformation.  This curve should be corrected by
  the thick dashed line.  The dotted line correspond to $q=0.5$, corrected by
  the bold dotted line.  The dotted-dashed line correspond to $q=0.55975$,
  where a reentrant behavior for a small range of t can still be noticed.  The
  thin line corresponds to results at $q=0.56$, where we do not need to
  correct the $\beta\rightarrow\beta'$ transformation.  }
\end{figure}

\begin{figure}[!ht]
\label{fig:ttl}
\epsfxsize=8cm \centerline{\rotate[r]{\epsfbox{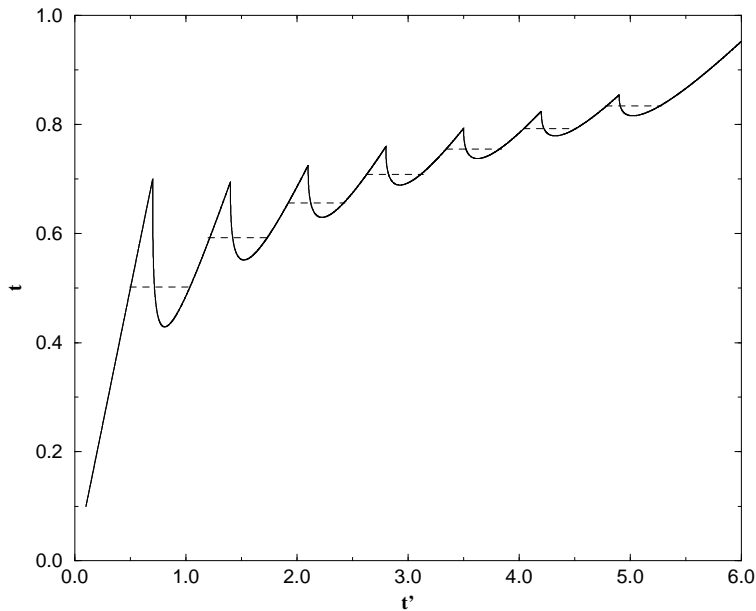}}}
\caption{
$t$ vs $t'$ for $N=8$ and $q=0.3$.  The dashed-line corresponds to the
cutting  points obtained from free  energy arguments.  We see that $t$
is an increasing function of $t'$. }
\end{figure}

Let us address now the role of the cut-off in the stability of the statistics.
In fig. 5 we present the free energy calculated with and without the cut-off
for $q=0.5$.  These calculations are only possible for $q=1-1/2n$ where
$n=1,2,3...$ since the exponent $1/(1-q)$ will be even.  As defined in the
step (1) of the $\beta\rightarrow\beta'$ recipe, $y_i$ might assume negative
values but because they are raised to an even power, $y_i^{1/(1-q)}$ will be
always real.  Without the cut-off step, the free energy shows an unstable
concave region.  When the cut-off is applied, the unstable region disappears.
It is clear that the unstable loop does not appear due to the cut-off.  For
these special values of $q$ we were able to calculate the roots of eq.
(\ref{piq}).  The reentrant region corresponds to three real and different
roots.  Outside the reentrant region, we obtain either one real and two
complex conjugated roots or a three-fold degenerate real ones.

\begin{figure}[!ht]
\label{fig:cut}
\epsfxsize=8cm \centerline{\rotate[r]{\epsfbox{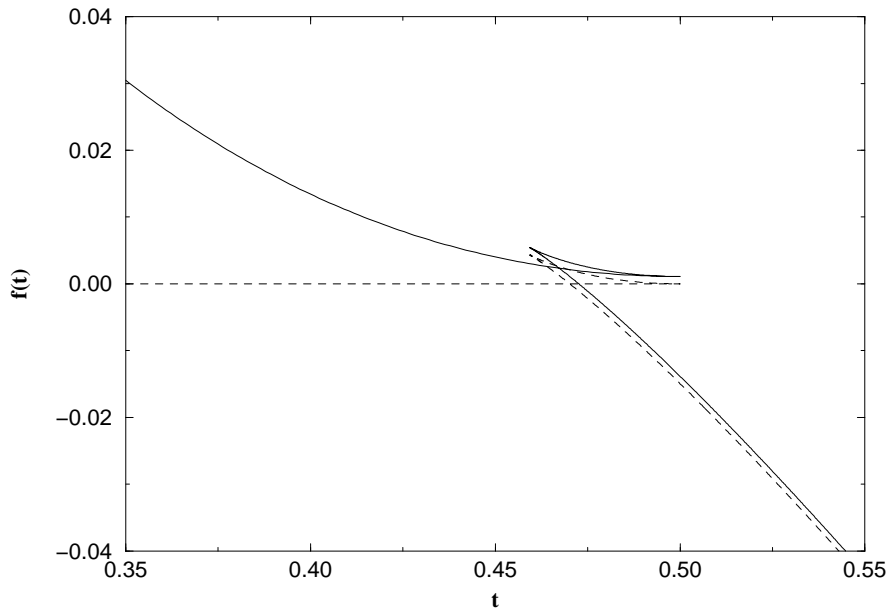}}}
\caption{Free energy for the system with (dashed lines) and without
  (continuous line) cut-off (see text).  Besides the loop, the solution
  without cut-off presents a concave region wich does not fill the stability
  requirements. The curves are slightly displaced up again for clarity.  }
\end{figure}

In summary, we have shown that the iterative procedure gives a non physical
discontinuity in the free energy.  An alternative procedure called ``$\beta
\rightarrow \beta'$ transformation'' must include a more careful treatment
based on free energy convexity arguments.  We show that the standard way of
choosing states with lowest free energy values allows us to get a correct
internal energy temperature dependency.  Since the reentrant behavior appears
from $\beta$ as function of $\beta'$, the present prescription restores the
proper behavior in the temperature dependence of the thermodynamic observable.
But the most important consequence is that the Tsallis statistics with
normalized q-expectation values is stable, in the sense that, for all $\beta$,
the requirements of convexity (concavity) for $q \leq 1$ ($q > 1$) in the free
energy are satisfied.  The current method has been also applied to the
two-dimensional Ising model in a non-extensive regime\cite{Ising2d}.

\section*{Acknowledgements}

This work was motivated from a discussion with J.  S. S\'a Martins, when we
faced the hard task of concluding ref.  \cite{Ising2d}.  In that problem we
deal with similar problems for the simulation of the Ising model in the
Tsallis statistics.  The authors are grateful to M.  S.  Figueira, C.
Moukarzel, A.  Scala and M.  Reza Sadr-Lahijany for useful discussions.  We
also are indebted to Professor C.  Tsallis for providing us a copy of
reference (1) prior to its publication. This work was partially supported by
CNPq and CAPES (Brazilian Agencies).

\end{document}